\documentclass[]{spie}  
\usepackage[]{graphicx}
\usepackage[]{subeqn,url}
\title{A spatiospectral localization approach to estimating 
potential fields on the surface of a sphere from noisy, incomplete
data\\ taken at satellite altitudes} 
\author{Frederik J.~Simons 
and F.~A.~Dahlen 
\skiplinehalf
Department of Geosciences, Princeton University, Guyot Hall,
Princeton, NJ, USA\\
}
\newcommand{\wigner}[6]
{\left(\!\barray{ccc} #1 & #2 & #3\\#4 & #5 & #6 \earray\!\right)}
\newcommand{\sfW}{{\textsf{\small W}}}

\newcommand{\domg}{\,d\Omega} 
\newcommand{\fracd}[2]{\frac{\displaystyle{#1}}{\displaystyle{#2}}} 
\newcommand{\intr}{\int_R}

\newcommand{\into}{\int_\Omega}

\newcommand{\also}{\quad\mbox{and}\quad}

\newcommand{\Ylm}{Y_{lm}} 
\newcommand{\Ylmrh}{Y_{lm}(\rhat)}

\newcommand{\Ylmp}{Y_{l'm'}} 
\newcommand{\Xlm}{X_{lm}}

\newcommand{\Xlpm}{X_{l'm}}

\newcommand{\Dlmlmp}{D_{lm,l'm'}}

\newcommand{\Dllp}{D_{ll'}}

\newcommand{\Drhrhp}{D(\rhat,\rhat')} 
 
\newcommand{\grhp}{g(\rhat')}

\newcommand{\Lpot}{(L+1)^2}

\newcommand{\suml}{\sum\limits}

\newcommand{\sumshL}{\suml_{l=0}^{L}\suml_{m=-l}^{l}}

\newcommand{\tlofp}{\left(\frac{2l+1}{4\pi}\right)} 
\newcommand{\Plm}{P_{lm}} 
 
\newcommand{\sfD}{{\textsf D}}

\newcommand{\rhat}{\mbf{\hat{r}}}

\newcommand{\be}{\begin{equation}} 
\newcommand{\ee}{\end{equation}} 
\newcommand{\ber}{\begin{eqnarray}} 
\newcommand{\eer}{\end{eqnarray}} 
\newcommand{\barray}{\begin{array}} 
\newcommand{\earray}{\end{array}} 
\newcommand{\mbf}{\mathbf}

\newcommand{\nnr}{\nonumber}

\newcommand{\rar}{\rightarrow}

\begin{document} 
\maketitle 

\begin{abstract}
Satellites mapping the spatial variations of the gravitational or
magnetic fields of the Earth or other planets ideally fly on polar
orbits, uniformly covering the entire globe. Thus, potential fields on
the sphere are usually expressed in spherical harmonics, basis
functions with global support. For various reasons, however, inclined
orbits are favorable. These leave a ``polar gap'': an antipodal pair
of axisymmetric polar caps without any data coverage, typically
smaller than 10$^\circ$ in diameter for terrestrial gravitational
problems, but 20$^\circ$ or more in some planetary magnetic
configurations. The estimation of spherical harmonic field
coefficients from an incompletely sampled sphere is prone to error,
since the spherical harmonics are not orthogonal over the partial
domain of the cut sphere. Although approaches based on wavelets have
gained in popularity in the last decade, we present a method for
localized spherical analysis that is firmly rooted in spherical
harmonics. We construct a basis of bandlimited spherical functions
that have the majority of their energy concentrated in a subdomain of
the unit sphere by solving Slepian's (1960) concentration problem in
spherical geometry, and use them for the geodetic problem at
hand. Most of this work has been published by us elsewhere. Here, we
highlight the connection of the ``spherical Slepian basis'' to
wavelets by showing their asymptotic self-similarity, and focus on the
computational considerations of calculating concentrated basis
functions on irregularly shaped domains.
\end{abstract}

\nocite{Slepian+60}


\section{INTRODUCTION}

Constructing local spherical harmonic bases that are orthogonal over
limited domains and still behave well under the action of up- and
downward continuation operators is of interest in geomagnetism
\cite{Thebault+2006,Lesur2006} and geodesy
\cite{Albertella+99,Simons+2006b}. As an alternative to a wavelet
basis~\cite{Holschneider+2003,Freeden+2004}, we construct a new basis
of so-called Slepian functions~\cite{Slepian83} on the sphere. These
bandlimited functions are designed to have the majority of their
energy optimally concentrated inside the geographically limited region
covered by satellites, as in Fig.~\ref{dsdiagram}. Slepian functions
are orthogonal on both the entire as well as the cut sphere, a
property that can be exploited to our
advantage. Elsewhere~\cite{Simons+2006b} we have studied the inverse
problem of retrieving a potential field on the unit sphere from noisy
and incomplete but continuously available observations made at an
altitude above their source.  We have obtained exact expressions for
the estimation error due to the traditional method of damped
least-squares spherical harmonic analysis as well as that arising from
a new approach which uses a truncated set of Slepian basis functions.

\begin{figure}[t]\centering 
\rotatebox{-90}{
\includegraphics[height=1\textwidth]
{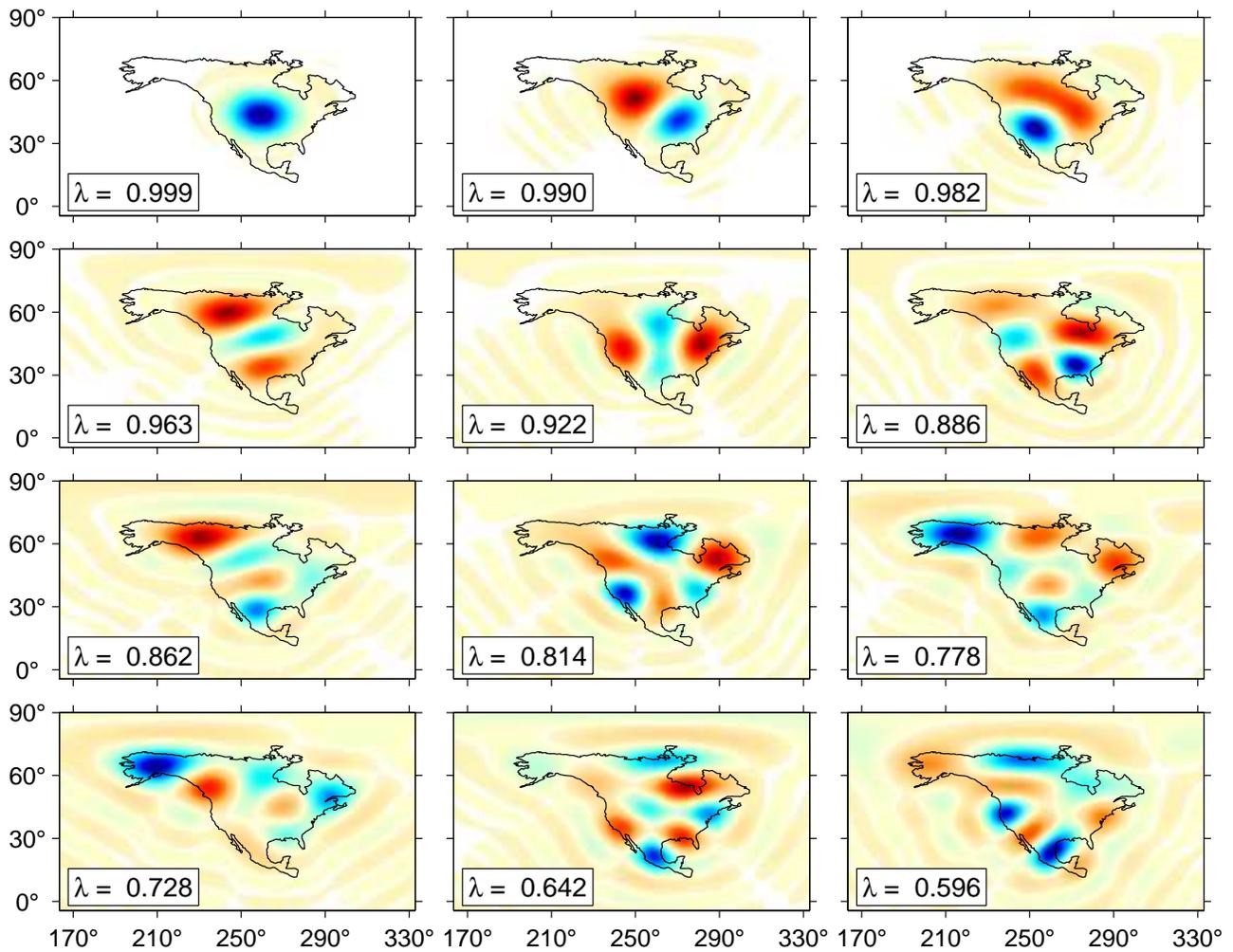}
} 
\caption{\small Bandlimited $L=18$ eigenfunctions
$g_1,g_2,\ldots,g_{12}$ that are optimally concentrated within the
continent of North America. The concentration factors
$\lambda_1,\lambda_2,\ldots,\lambda_{12}$ are indicated; the Shannon
number is $N=14$.}
\label{namerica} 
\end{figure}

\begin{figure}[t]\centering 
\rotatebox{-90}{
\includegraphics[width=0.925\textwidth]
{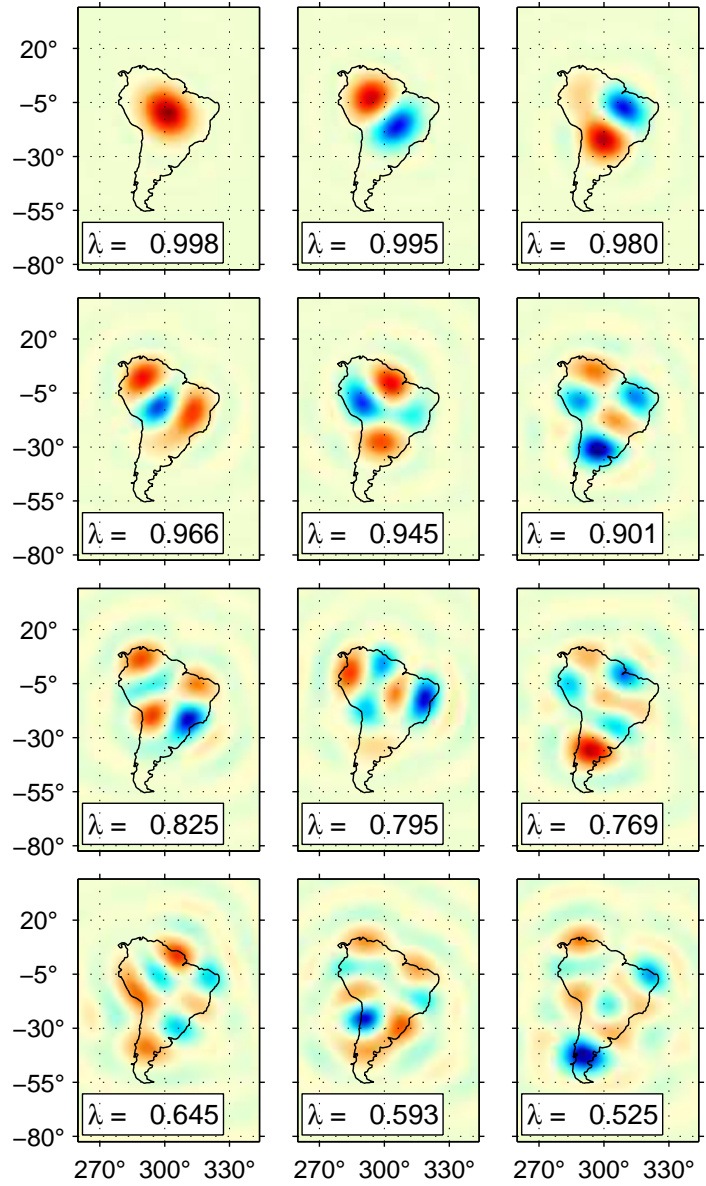}
} 
\caption{\small Bandlimited $L=18$ eigenfunctions
$g_1,g_2,\ldots,g_{12}$ that are optimally concentrated within the
continent of South America. The Shannon number is $N=13$. Format is identical to
that in Fig.~\ref{namerica}.}
\label{samerica} 
\end{figure}

\begin{figure}\centering 
\rotatebox{-90}{
{\includegraphics[height=0.725\textwidth]
{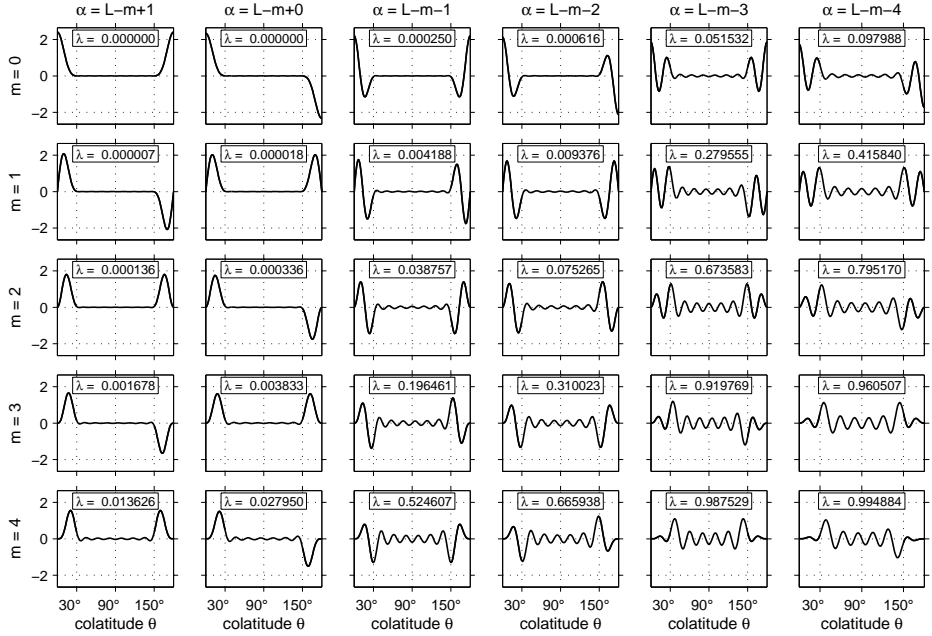}}
} 
\caption{Colatitudinal dependence of the last six fixed-order,
$m=0\rar4$, eigenfunctions $g_\alpha(\theta)$, $\alpha=L-m+1\rar
L-m-4$, bandlimited to $L=18$. These are generally poorly concentrated
in the latitudinal belt $\pm60^{\circ}$ about the equator, except
where the rank $\alpha$ exceeds the fixed-order Shannon number $N_m$
(examples in lower right). The functions that have the least energy
inside of the equatorial belt, as shown by their low eigenvalues
$\lambda_\alpha$, are best concentrated inside the complementary polar
caps of colatitudinal radius $\Theta=30^{\circ}$.}
\label{sdspace}  
\end{figure}

\begin{figure}\centering 
\rotatebox{-90}{
{\includegraphics[height=0.725\textwidth]
{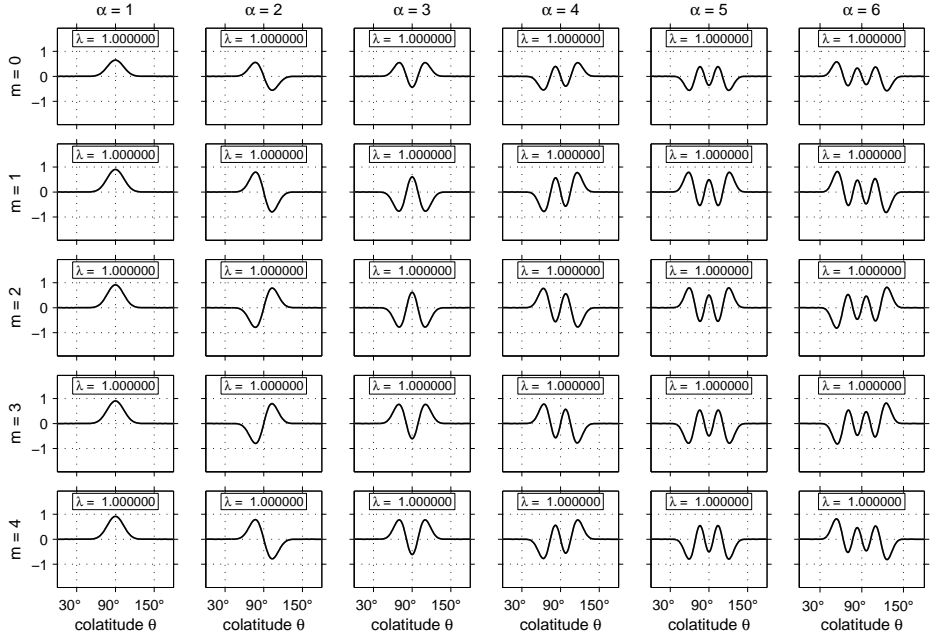}}  
} 
\caption{Colatitudinal dependence of the first six fixed-order,
$m=0\rar4$, eigenfunctions $g_\alpha(\theta)$, $\alpha=1\rar6$,
bandlimited to $L=18$, that are well concentrated in the latitudinal
belt extending $\pm60^{\circ}$ on either side of the equator. The
quality of the spatial concentration is expressed by the labeled
eigenvalues $\lambda_\alpha$. None of the plotted functions show
appreciable energy inside the complementary pair of antipodal polar
caps of radius $\Theta=30^{\circ}$.}
\label{sdbelt}
\end{figure}

The geodetic estimation problem can be cast in the much wider context
of spatiospectral localization, whereby bandlimited functions are
spatially concentrated to regions of arbitrary shape on the sphere
\cite{Wieczorek+2005,Simons+2006a}. Some of these are illustrated in
Figs~\ref{namerica}--\ref{samerica}. A semi-analytical numerical
method can be used to calculate the spherical Slepian functions on a
latitudinal belt symmetric about the equator, or its complement, the
double polar cap. This approach requires no numerical integration and
avoids the construction of matrices other than a tridiagonal matrix
whose elements are prescribed analytically. Finding spherical harmonic
expressions for bandlimited functions concentrated to polar caps, as
in Fig.~\ref{sdspace}, or latitudinal belts, as in Fig.~\ref{sdbelt},
thus becomes so effortless as to be achievable by a handful of lines
of computer code, and the problems with numerical stability that are
known to plague alternative approaches~\cite{Albertella+99,Pail+2001}
are avoided altogether.  The key to this ``magic'' lay hidden in two
little-known studies published several decades ago: the work by
Gilbert~\cite{Gilbert+77} on doubly orthogonal polynomials, and that
on commuting differential operators by Gr\"unbaum
\cite{Grunbaum+82}. It must be remembered that one of Slepian's main
discoveries~\cite{Slepian83} was the existence of a second-order
differential operator that commutes with the spatiospectral
localization kernel concentrating to intervals on the real
line. Finding the ``prolate spheroidal functions'' amounts to the
diagonalization of a simple tridiagonal matrix
\cite{Percival+93}. Gilbert~\cite{Gilbert+77} presented two additional
commuting differential operators, which are applicable to the
concentration of Legendre polynomials to one-and two-sided domains.
Gr\"unbaum~\cite{Grunbaum+82} proved that the matrix accompanying the
localization to the single polar cap is, once again, tridiagonal, and
the same holds for the double polar cap, or its complement, the
latitudinal belt, as we have shown~\cite{Simons+2006b}.

\begin{figure}\center
\rotatebox{0}{
\includegraphics[width=0.61\textwidth]{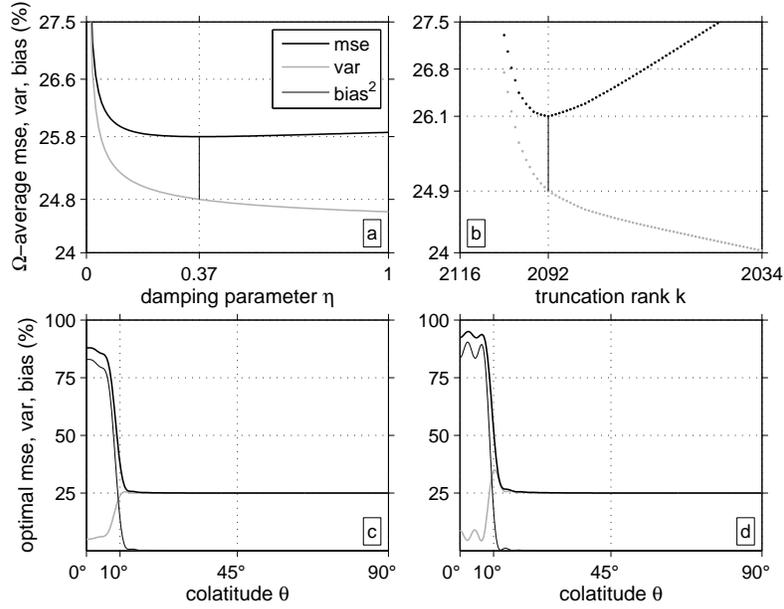}
}
\caption{Mean-squared error (mse), variance and bias for the damped
least-squares solution to the geodetic inverse problem of recovering
the source signal from incomplete and noisy observations, and the
truncated Slepian approach~\cite{Simons+2006b}. The antipodal polar
gap has a radius $\Theta=10^{\circ}$; the bandwidth is $L=45$;
signal-to-noise ratio $S/N=4$. Panels a \& b show the values averaged
over the unit sphere; the squared bias is the difference between the
mse and variance curves, indicated by the thin black vertical
line. Panels c \& d show the values at the optimal damping and
truncation levels.}
\label{sdhoerl}  
\end{figure}

In practice, the geodetic and geomagnetic inverse problems are always
ill-conditioned, even in the absence of a polar gap, due the
peculiarities of orbital data coverage and the distribution of noise
sources~\cite{Xu1992b,Holme+95}. In the standard method of damped
spherical harmonic inversion, the ill-conditioning is alleviated by
the addition of a small damping parameter to the normal equation
matrix prior to inversion. Often the value of this parameter is {\it
ad hoc} and chosen primarily for numerical stability, but more
sophisticated methods use a priori statistical information about the
set of model parameters. We have derived the exact structure of the
model parameter sensitivity matrix arising from the presence of a
contiguous data gap, assuming the data are known continuously
everywhere but inside it. The Slepian functions were revealed to be
the very eigenfunctions of this matrix. Assuming a particular
covariance structure for the model parameters and the observational
noise, this knowledge allowed us to write analytical expressions for
the optimal regularization terms for the damped spherical harmonic
method. Such an approach optimally filters out the small eigenvalues,
and thus reduces the ill-conditioning of the sensitivity matrix
\cite{Mallat98}. Our preferred method~\cite{Simons+2006b} applies a
hard truncation to the singular values of the sensitivity matrix in an
approach based directly on the Slepian expansion of the model. We have
showed that this is only marginally less successful in minimizing the
mean-squared estimation error, as shown in Fig.~\ref{sdhoerl} for the
case of white signal and noise, as well as computationally
advantageous and more intuitively appealing.

The problems we posed and solved in this context are not limited to
geodesy and observations made from a satellite. In geomagnetism, our
observation level may be the Earth's surface, and the source level at
or near the core-mantle boundary~\cite{Lowes74,Gubbins83}. In
cosmology, the unit sphere constituting the sky is observed from the
inside out, and the galactic plane masking spacecraft measurements has
the shape of a latitudinal belt~\cite{Tegmark96a}. Ground-based
astronomical measurements may be confined to a small circular patch of
the sky~\cite{Peebles73,Tegmark95}. Finally, in planetary science,
knowledge of the estimation statistics of properties observed over
mere portions of the planetary surface is important in the absence of
groundtruthing observations.

\section{Concentration within an arbitrarily shaped region} 
\label{sec:Arbitrary} 

We seek to determine those bandlimited functions $g(\rhat)$ that
are optimally concentrated within a spatial region~$R$.

\subsection{Spherical harmonics}

The geometry of the unit sphere $\Omega=\{\rhat: \|\rhat\|=1\}$ is
depicted in Fig.~\ref{dsdiagram}.  We denote the colatitude of a
geographical point $\rhat$ by $0\le\theta\le\pi$ and the longitude by
$0\le\phi< 2\pi$; the geodesic angular distance between two points
$\rhat$ and $\rhat'$ will be denoted by $\Delta$. We use $R$ to denote
a region of $\Omega$, of area $A$, within which we seek to concentrate
a bandlimited function of position $\rhat$.  We use real surface
spherical harmonics defined
by~\cite{Dahlen+98,Edmonds96} 
\ber
\Ylmrh= \Ylm(\theta,\phi)&=&\left\{
\begin{array}{l@{\quad\mbox{if}\hspace{0.6em}}l}
\rule[-2mm]{0mm}{6mm}\sqrt{2}X_{lm}(\theta)\cos m\phi & -l\le m<0\\
\rule[-2mm]{0mm}{6mm}X_{l0}(\theta)                     & m=0\\
\rule[-2mm]{0mm}{6mm}\sqrt{2}\Xlm(\theta)\sin m\phi & 0< m\le l,\\
\end{array}
\right.\label{Ylm}\\
\Xlm(\theta)&=&
(-1)^m\tlofp^{1/2}
\left[\frac{(l-m)!}{(l+m)!}\right]^{1/2}\!\Plm (\cos\theta),
\label{xlm}\\
\Plm (\mu)&=&
\frac{1}{2^ll!}(1-\mu^2)^{m/2}\left(\frac{d}{d\mu}\right)^{l+m}\!(\mu^2-1)^l.
\label{plm}
\eer The quantity $0\le l\le\infty$ is the angular degree of the
spherical harmonic, and $-l\le m\le l$ is its angular order. The
function $\Plm(\mu)$ defined in (\ref{plm}) is the associated Legendre
function of integer degree $l$ and order $m$. Our choice of the constants in
equations~(\ref{Ylm})--(\ref{xlm}) orthonormalizes the harmonics on
the unit sphere:
\begin{equation}
\into\Ylm\Ylmp\domg=\delta_{ll'}\delta_{mm'}.
\label{normalization}
\end{equation}

\subsection{Spatial concentration of a bandlimited function to an
  arbitrarily shaped region}
To maximize the spatial concentration of a bandlimited
function
\be
g=\sumshL g_{lm}Y_{lm},
\label{bandlg}
\ee
within a region $R$,
we maximize the ratio
\begin{equation}
\lambda=\intr g^2\domg\left/\into^{}g^2\domg\right.
.
\label{normratio}
\end{equation}
Elsewhere, we have shown~\cite{Simons+2006a} that maximizing
equation~(\ref{normratio}) leads to the eigenvalue equation  
\begin{equation}
\suml_{l'=0}^L\suml_{m'=-l'}^{l'}\Dlmlmp
g_{l'm'}=\lambda\hspace{0.05em}g_{lm},\quad
\mbox{where}
\quad\Dlmlmp=\intr\Ylm\Ylmp\domg.
\label{fulleigen1}
\label{Dlmlmpdef}
\end{equation}
We have also shown~\cite{Simons+2006a}
that we may rewrite equation~(\ref{fulleigen1}) as a spatial-domain
eigenvalue equation: 
\begin{equation}
\intr  \Drhrhp \,\grhp\domg'=
\lambda\hspace{0.05em}g(\rhat),
\quad\rhat\in\Omega,
\quad
\mbox{where}
\quad
\Drhrhp
=\sum_{l=0}^L\left(\frac{2l+1}{4\pi}\right)\!P_l(\cos\Delta).
\label{banddelta}
\label{firsttimeint}
\end{equation}
Equation~(\ref{firsttimeint}) is a homogeneous Fredholm
integral equation of the second kind, with a finite-rank, symmetric,
separable kernel~\cite{Kanwal71,Tricomi70}.
The spectral-domain eigenvalue problem~(\ref{fulleigen1}) for the
spherical harmonic expansion coefficients~$g_{lm}$ and the
spatial-domain eigenvalue problem~(\ref{firsttimeint}) for the
spatial-domain $g(\rhat)$ are completely equivalent. The sum of the
eigenvalues of equations~(\ref{fulleigen1}) or~(\ref{firsttimeint}) is
given by 
\be
\qquad N=\sum_{\alpha =1}^{\Lpot}\lambda_{\alpha}=
\sum_{l=0}^L\sum_{m=-l}^l 
D_{lm,lm}= \int_RD(\rhat,\rhat)\,d\Omega
=\Lpot\,\frac{A}{4\pi}
.
\label{tracedef}
\ee 
This is the spherical analogue of the ``Shannon number'' in Slepian's
one-dimensional concentration problem
\cite{Slepian+60,Slepian83,Percival+93}.  

\subsection{Spatial concentration of a bandlimited function to
azimuthally symmetric regions}

\begin{figure}\center
\rotatebox{0}{
\includegraphics[width=\columnwidth]{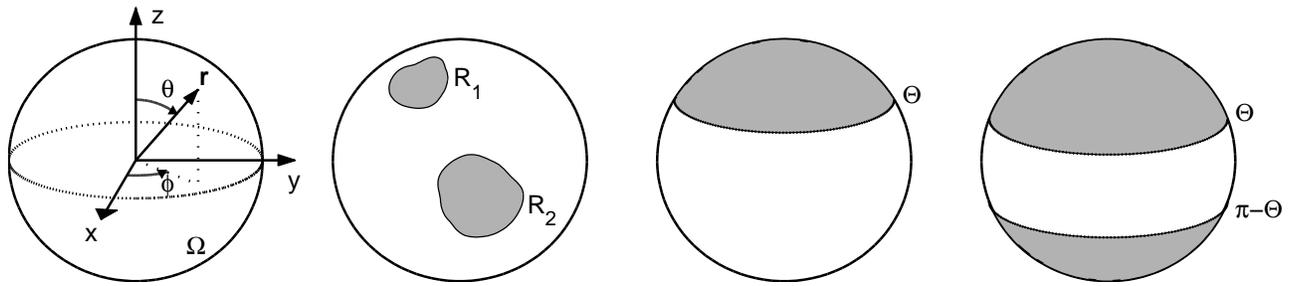}
}
\caption{Geometry of the geodetic estimation problem: a concentration
region of arbitrary geometry; an axisymmetric polar cap, shaded, of
colatitudinal radius $\Theta$; an antipodal pair of polar caps, shaded,
representing the geodetic polar gap.}
\label{dsdiagram}  
\end{figure}

\begin{figure}[b]\centering 
\rotatebox{0}{
\includegraphics[width=0.61\textwidth]{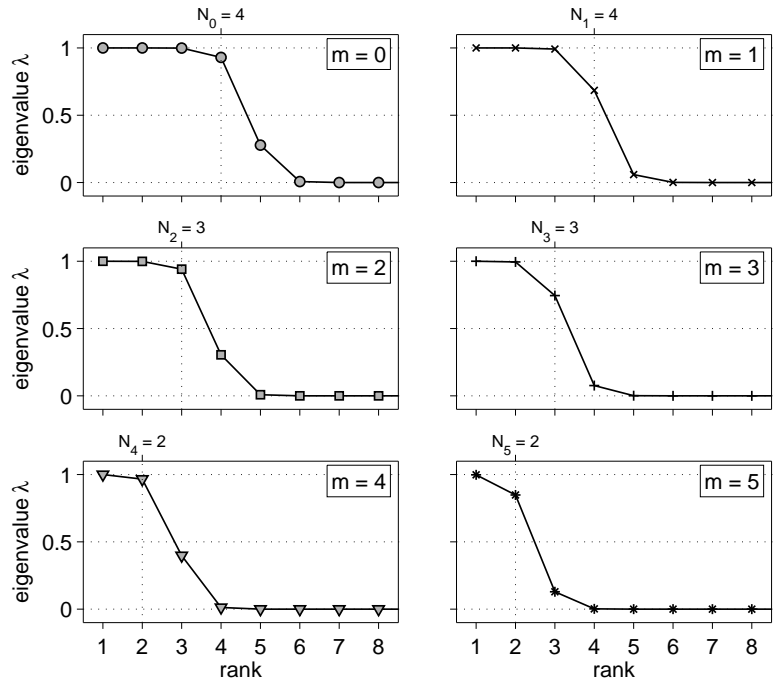}
} 
\caption{\small Fixed-order eigenvalue spectra for an axisymmetric
polar cap of radius $\Theta=40^{\circ}$.  The maximal spherical
harmonic degree is $L=18$.  A different symbol is used to plot
$\lambda_{\alpha}$ versus rank $\alpha$ for each order $0\leq m\leq
5$. The total number of fixed-order eigenvalues is $L-m+1$; only the
largest eight ($\lambda_1$ through $\lambda_8$) are shown.  Vertical
grid lines and top labels specify the partial Shannon numbers $N_m$,
rounded to the nearest integer.}
\label{sdweigen} 
\end{figure}

In the important special case in which the region of concentration
is a circularly symmetric cap of colatitudinal radius $\Theta$,
centered on the north pole, as shown in Fig.~\ref{dsdiagram},
the matrix elements~(\ref{Dlmlmpdef}) reduce
to~\cite{Simons+2006a,Simons+2006b}  
\ber
\Dllp &=&
2\pi\,\int_{0}^{\Theta}
\Xlm \Xlpm\sin\theta\,d\theta
\quad\mbox{for each order}\quad
0\le|m|\le L, 
\label{blockdia}\\
&=&(-1)^m\frac{\sqrt{(2l+1)(2l'+1)}}{2}\!\!\sum_{n=|l-l'|}^{l+l'}
\wigner{l}{n}{l'}{0}{0}{0}
\wigner{l}{n}{l'}{m}{0}{\!\!-m}
\big[P_{n-1}(\cos\Theta)-P_{n+1}(\cos\Theta)\big],
\label{kernel3j}
\eer
where the arrays of indices are Wigner 3-$j$ symbols
\cite{Edmonds96,Dahlen+98,Messiah2000}. 
For a circularly symmetric double cap
of common colatitudinal radius $\Theta$, as in
Fig.~\ref{dsdiagram}, we obtain~\cite{Simons+2006b}
\be
D_{ll'}=2\pi\left[1+(-1)^{l+l'}\right]
\int_{0}^{\Theta}\Xlm\Xlpm\sin\theta\,d\theta
\quad\mbox{for each order}\quad
0\le|m|\le L.
\label{kernel5}
\ee
The double-cap spatial eigenfunctions separate into solutions that
are even or odd across the equator~\cite{Simons+2006a}. Finally, the
single-order equivalent of the spatial-domain
kernel~(\ref{firsttimeint}) is given by~\cite{Simons+2006a} \be
D(\mu,\mu')=\fracd{(L-m+1)!}{2(L+m)!}
\left[\fracd{P_{L+1\,m}(\mu)P_{Lm}(\mu')-P_{Lm}(\mu)P_{L+1\,m}(\mu')}
{\mu-\mu'}\right],
\label{CDkernel}
\ee
where $\mu=\cos(\theta)$ and which can be computed using
L'H\^{o}pital's rule when $\mu=\mu'$. For each of the fixed-order
eigenvalue problems the partial Shannon number can be computed from
\be
N_m=\sum_{\alpha=1}^{L-m+1}\!\!\lambda_{\alpha} =\suml_{l=m}^{L}
D_{ll} =\int_{\cos\Theta}^{1}D(\mu,\mu)\,d\mu
=\fracd{(L-m+1)!}{2(L+m)!}\int_{\cos\Theta}^1 
\big[P_{L+1\,m}^{\prime}P_{Lm}-P_{Lm}^{\prime}P_{L+1\,m}\big]\,d\mu,
\label{nsubm}\label{nsubm2}
\ee
where the prime denotes differentiation with respect to
$\mu$. Once the $L+1$ sequences of fixed-order eigenvalues have been found,
they can be resorted to exhibit an overall mixed-order ranking. The
total number of significant eigenvalues (\ref{tracedef}) is then
\be
N=N_0+2\sum_{m=1}^{L}N_m,\label{Nequalsum}
\ee
where the factor of two accounts for the $\pm m$ degeneracy.
Fig.~\ref{sdweigen} shows the fixed-order eigenvalue spectra for
$0\leq m\leq 5$. The cap radius is $\Theta=40^{\circ}$ and the maximal
spherical harmonic degree is $L=18$. The partial Shannon numbers
$N_m$, computed by rounding equation~(\ref{nsubm}) to the nearest
integer, are shown. As in the case of the classical Slepian problem
\cite{Landau65,Percival+93,Slepian+65} the spectra have a
characteristic step shape, showing significant ($\lambda\approx 1$)
and insignificant ($\lambda\approx 0$) eigenvalues separated by a
narrow transition band. The partial Shannon number~(\ref{nsubm})
provides a good estimate of the number of well concentrated
eigenfunctions; the first $N_m$ eigenfunctions all have a
concentration factor exceeding $\lambda=0.5$.

As we showed elsewhere~\cite{Simons+2006a,Simons+2006b},
equation~(\ref{fulleigen1}) for the single cap is equivalent to an
algebraic eigenvalue equation that requires diagonalization of a
matrix that commutes with~(\ref{blockdia})--(\ref{kernel3j}) and whose
elements are given by 
\ber T_{ll}&=&-l(l+1)\cos{\Theta},\nnr\\
T_{l\,l+1}&=&\big[l(l+2)-L(L+2)\big]
\sqrt{\fracd{(l+1)^2-m^2}{(2l+1)(2l+3)}},
\label{gdefi1}
\eer
and we can replace the matrix~(\ref{kernel5}) for 
functions odd ($p=o$) or even ($p=e$)  across the equator by 
the matrix
\ber
T^p_{ll}&=&-l(l+1)\cos^2{\Theta}
+\frac{2}{2l+3}\left[(l+1)^2-m^2\right]
+[(l-2)(l+1)-L_p(L_p+3)]
\left[\frac{1}{3}-\frac{2}{3}\,
\fracd{3m^2-l(l+1)}{(2l+3)(2l-1)}\right],\nnr\\ 
T^p_{l\,l+2}&=&\fracd{\big[l(l+3)-L_p(L_p+3)\big]}{2l+3}
\sqrt{\fracd{\left[(l+2)^2-m^2\right]
\left[(l+1)^2-m^2\right]}{(2l+5)(2l+1)}}.
\label{gdefi}
\eer
for the double cap, where $L_e$ and $L_o$ are $L$ and $L-1$,
respectively, if $m$ and $L$ have the same parity, and $L-1$ and $L$,
respectively, if $m$ and $L$ have opposite parity. Thus,
equation~(\ref{gdefi}) defines two matrices, one for the even and one
for the odd functions; both are properly speaking
tridiagonal~\cite{Simons+2006b}. 

\section{Asymptotic scaling}
\label{sec:Asymptotics}

The eigenvalues $\lambda_1,\lambda_2,\ldots$
and suitably scaled eigenfunctions $\psi_1(x),\psi(x)_2,\ldots$
of Slepian's time series problem~\cite{Slepian+60}, of
optimally concentrating a strictly bandlimited signal with a
spectrum that vanishes for frequencies $|\omega|>W$ into
a time interval $|t|\le T$, in other words, the eigenvalues and
eigenfunctions of  
\be
\int_{-1}^{1}\fracd{\sin TW(x-x')}
{\pi (x-x')}\,\psi(x')\,dx'=
\lambda\hspace{0.05em}\psi(x),\quad |x|\le 1,
\label{slepian5}
\ee
depend only upon the ``Shannon number'' $N=2TW/\pi$. This scaling is
the only important feature of the one-dimensional problem that does
not carry over to the spatiospectral concentration problem on a
sphere.  Shannon-number scaling on a sphere is exhibited only
asymptotically, in the limit
\begin{equation}
A\rightarrow 0,\quad L\rightarrow\infty,\quad
\mbox{with}\quad
N=\Lpot\fracd{A}{4\pi}
\quad\mbox{held fixed}. 
\label{asylimit}
\end{equation}
In that limit of a small concentration area $A$ and a
large bandwidth $0\le l\le L$, the curvature of the
sphere becomes negligible and the spherical
concentration problem approaches the concentration problem
in the plane~\cite{Slepian64}. 

\subsection{Scaled integral equation for an arbitrarily shaped region}

Two  results underlie the consideration of the flat-Earth
limit~(\ref{asylimit}), which we undertake in this section.
The first is Hilb's asymptotic
approximation for the Legendre functions
\cite{Amado+85,Dahlen80,Hilb19,Szego75}, 
\begin{equation}
X_{lm}(\theta)\approx (-1)^m\sqrt{\frac{l+1/2}{2\pi}}\,
\sqrt{\frac{\theta}{\sin\theta}}\,
J_m\big[(l+1/2)\theta\big],\quad 0\le\theta\ll\pi,
\label{Hilb}
\end{equation}
where $J_m(x)$ is the Bessel function of the first kind;
the second is the truncated Watson-Poisson sum
formula~\cite{Dahlen+98}, 
\begin{equation}
\suml_{l=0}^L{f}(l+1/2)=\sum_{s=-\infty}^{\infty}(-1)^s
\int_{0}^{L+1}{f}(k)e^{-2\pi isk}\,dk,
\label{Poisson}
\end{equation}
valid for an arbitrary continuous function $f(x)$. An
application of~(\ref{Hilb}) and~(\ref{Poisson}), substituting
$k=(L+1)p$ and taking the limit $L\rightarrow\infty,\Delta\rightarrow
0$, with the product $L\Delta$ held fixed, 
enables us to write the Fredholm kernel $\Drhrhp$ of
equation~(\ref{banddelta}) in the form  
\begin{equation}
D(\Delta)\approx\frac{\Lpot}{2\pi}
\int_{0}^{1}J_0\big[(L+1)p\Delta\big]\,p\,dp\nnr
=\frac{(L+1)\,J_1\big[(L+1)\Delta\big]}{2\pi\Delta},
\label{smallK}
\end{equation}
where we have made the approximation $\Delta/\!\sin\Delta\approx 1$,
and used the Riemann-Lebesgue lemma~\cite{Olver97} to eliminate the
$s\not= 0$ terms involving the highly oscillatory factors $e^{-2\pi
is(L+1)p}$.  In the limit $x\rightarrow 0$ the ratio
$J_1(x)/x\rightarrow 1/2$, so the $\Delta\rightarrow 0$ limit
of~(\ref{smallK}) is $D(0)=\Lpot/(4\pi)$, guaranteeing that  
the Shannon number remains unchanged.

To obtain a scaled version of equation~(\ref{firsttimeint})
dependent only upon the Shannon number $N$, we make use of the
approximation~(\ref{smallK}) for the kernel $D(\rhat,\rhat')$,
and introduce the independent and dependent variable transformations
\begin{equation}
\label{hugesphere}
{\bf x}=\sqrt{\fracd{4\pi}{A}}\,\rhat,\quad
{\bf x}'=\sqrt{\fracd{4\pi}{A}}\,\rhat',\qquad
\psi({\bf x})=g(\rhat),\quad
\psi({\bf x}')=g(\rhat').
\end{equation}
The scaled coordinates ${\bf x},{\bf x}'$ are the
projections of the points $\rhat,\rhat'\in\Omega$ onto a large
sphere $\Omega_{*}$ of squared radius $\|{\bf x}\|^2=4\pi/A$.
The geodesic distance
between the points ${\bf x},{\bf x}'\in\Omega_{*}$
and the differential surface area on $\Omega_{*}$ are
\begin{equation}
\|{\bf x}-{\bf x}'\|=\sqrt{\fracd{4\pi}{A}}\,\Delta\quad
\mbox{and}\quad d\Omega_{*}=\frac{4\pi}{A}\,d\Omega.
\label{Deltatilde}
\end{equation}
Upon making the substitutions~(\ref{hugesphere})--(\ref{Deltatilde}),
equations~(\ref{firsttimeint}) and~(\ref{smallK}) reduce to
\begin{equation}
\label{scaledup}
\int_{R_{*}}\!\!D_{*}({\bf x},{\bf x}')\,\psi({\bf x'})\,d\Omega_{*}'
=\lambda\hspace{0.1em}\psi({\bf x}),\quad\mbox{with}\quad
D_{*}({\bf x},{\bf x}')=\fracd{\sqrt{N}}{2\pi}\,
\fracd{J_1\big(\raisebox{-0.2ex}{$\sqrt{N}$}\,\|{\bf x}-{\bf x}'\|\big)}
{\|{\bf x}-{\bf x}'\|},
\end{equation}
where $R_{*}$, of area $4\pi$, is the projection of the region of
concentration $R$ onto the sphere $\Omega_{*}$ and $D_{*}({\bf x},{\bf
x}')$ is the symmetric, $N$-dependent Fredholm
kernel. Equation~(\ref{scaledup}) is the spherical analogue of the
one-dimensional scaled eigenvalue equation~(\ref{slepian5}).  The
asymptotic eigenvalues $\lambda_1,\lambda_2,\ldots$ and associated
scaled eigenfunctions $\psi_1({\bf x}),\psi_2({\bf x}),\ldots$ depend
upon the maximal degree $L$ and the area $A$ only through the Shannon
number. As in the case of equation~(\ref{firsttimeint}), we are free
to solve~(\ref{scaledup}) either on all of
$\Omega_{*}$, in which case the eigenfunctions $\psi_1({\bf
x}),\psi_2({\bf x}),\ldots$ are bandlimited, or only in the region of
concentration $R_{*}$, in which case they are spacelimited.  It is
readily verified that the scaling has no effect upon the sum of the
eigenvalues.

\begin{figure}[b]\centering 
\rotatebox{0}{
\includegraphics[width=0.65\textwidth]{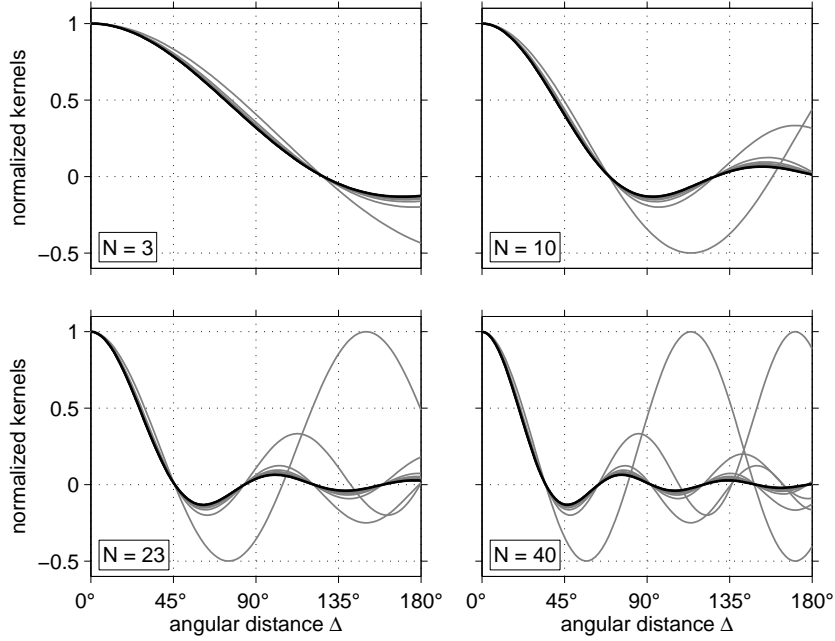}
} 
\caption{\small Comparison of the exact scaled kernels~(\ref{exactscaled})
with the flat-Earth asymptotic approximation~(\ref{asympscaled}) (black).
The Shannon number $N=3,10,23,40$ is kept constant in
each of the four panels, and the bandwidth used to compute the
exact scaled kernels varies between
$L=1$ (worst fitting) and $L=100$ (best fitting).}
\label{sdwscaling} 
\end{figure}

In the limit~(\ref{asylimit}), we expect the exact Fredholm
kernel~(\ref{banddelta}), evaluated on~$\Omega_{*}$ and normalized by
its value at zero offset,
\begin{equation}
\fracd{D\big(\sqrt{4\pi/A}\,\Delta\big)}{D(0)}=
\frac{1}{\Lpot}\suml_{l=0}^{L}(2l+1)\,
P_l\!\left(\cos\sqrt{\frac{4\pi}{A}}\,\Delta\right),
\label{exactscaled}
\end{equation}
to be well approximated by the similarly normalized asymptotic kernel
\begin{equation}
\fracd{D_{*}(\Delta)}{D_{*}(0)}=
\frac{2J_1(\sqrt{N}\Delta)}{\sqrt{N}\Delta}.
\label{asympscaled}
\end{equation}
The quality of this asymptotic approximation to the kernel
and the associated flat-Earth scaling are illustrated
in Fig.~\ref{sdwscaling}. In the four examples shown,
with Shannon numbers $N=3,10,23,40$, the approximation
is excellent even for angular distances as large as $\Delta\approx 135^{\circ}$,
once the maximal spherical harmonic degree exceeds $L=3$\,--\,$4$.

\subsection{Asymptotic fixed-order Shannon number}
\begin{figure}[h]\centering 
\rotatebox{0}{
\includegraphics[width=0.65\textwidth]{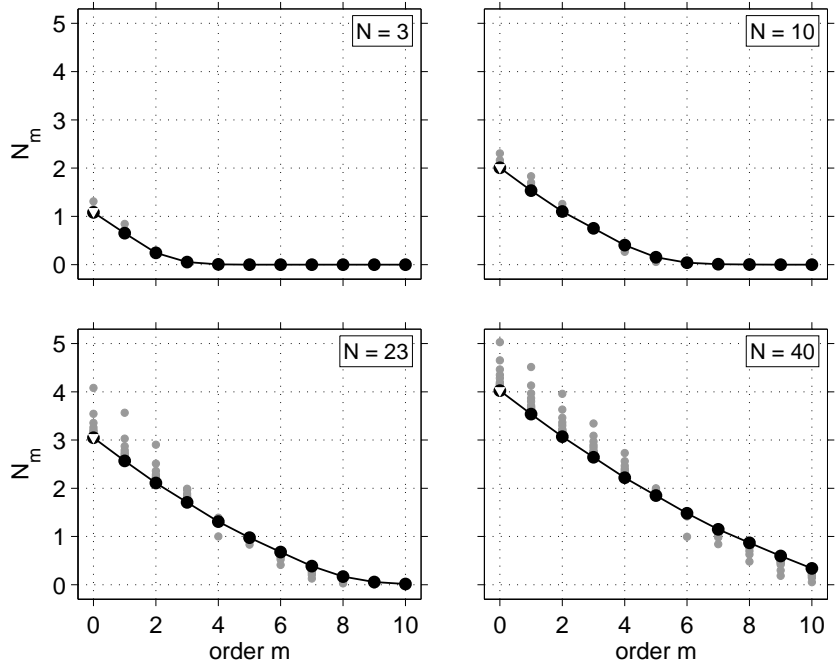}
} 
\caption{\small Comparison of the number $N_m$ of significant
eigenvalues of fixed order $m$ (gray) with the asymptotic
approximation~(\ref{nsubm3}) (black).  The Shannon number
$N=3,10,23,40$ is kept constant in each of the four panels, and the
bandwidth used to compute the exact values of $N_m$ varies between
$L=1$ (worst fitting) and $L=100$ (best fitting). Points inconsistent
with the constraint $A/(4\pi)=N/\Lpot <1$ are not plotted. White
triangles show the simplified zonal ($m=0$) approximation $N_0\approx
(L+1)\Theta/\pi$.}
\label{sdwnsubm} 
\end{figure}

The asymptotic approximation to the number of significant
eigenvalues associated with a given order $m$ is
\ber
N_m&=&\int_0^1D_{*}(x,x)\,x\,dx=4N\int_0^1\!\!\int_0^1\!
J_m^2\big(2\sqrt{N}\,p\hspace{0.1em}x\big)\,p\,dp\,x\,dx\nnr\\
&=&N^{m+1}{}_2F_3(1+m,1/2+m;1+2m,2+m,2+m;-4N)/\left[\Gamma(2+m)\right]^2\nnr\\ 
&=&2N\left[J^2_m(2\sqrt{N})+J^2_{m+1}(2\sqrt{N})\right]
-(2m+1)\sqrt{N}J_m(2\sqrt{N})J_{m+1}(2\sqrt{N})\nnr\\
&&{}-\frac{m}{2}\Big[1-J_0^2(2\sqrt{N})-2\suml_{n=1}^{m}J_n^2(2\sqrt{N})
\Big],
\label{nsubm3}
\eer where $F$ is a generalized hypergeometric function and $\Gamma$
the gamma function. The 
relationship~(\ref{Nequalsum}) between the total number $N$ of
significant eigenvalues and the number $N_m$ associated with each
order $m$ is preserved in this asymptotic approximation, inasmuch as,
by virtue of the identity $J_0^2(x)+2\sum_{m=1}^{\infty}J_m^2(x)=1$, \be
N=4N\int_{0}^{1}\!\!\int_{0}^{1} \left[J_0^2\big(2\sqrt{N}pq\big)+
2\suml_{m=1}^{\infty}J^2_m\big(2\sqrt{N}pq\big) \right]\,p\,dp\,x\,dx
=4N\int_{0}^{1}\!\!\int_{0}^{1}\,p\,dp\,x\,dx=N.  \ee In
Fig.~\ref{sdwnsubm} we compare the exact fixed-order Shannon numbers
$N_m$, computed by Gauss-Legendre numerical integration of
equation~(\ref{nsubm2}), with the asymptotic result~(\ref{nsubm3}),
for the same values of $N=3,10,23,40$ and $1\leq L\leq 100$ as in
Fig.~\ref{sdwscaling}.  The number of significant $m=0$ eigenvalues
can be even more simply approximated by $N_0\approx
2\sqrt{N}/\pi\approx (L+1)\Theta/\pi$, as shown. This can be derived
using the large-argument asymptotic expansion of the Bessel function
\cite{Jeffreys+88,Olver97}. The result~(\ref{nsubm3}) is
exact in the case of concentration in a two-dimensional plane.

\section{Computational considerations}
\label{sec:Computational}

All of the computations described here have been performed using
double precision arithmetic. Statements regarding machine precision
refer to double precision, with a round-off error of $\sim\! 10^{-16}$.

\subsection{Concentration within a polar cap}
\label{subsec:Comp:polar}
We may compute the colatitudinal eigenfunctions
$g_1(\theta),g_2(\theta),\ldots,g_{L-m+1}(\theta)$ of an axisymmetric
polar cap with $0\le\theta\le\Theta$ using three different methods. The
first is by numerical diagonalization of the
$(L-m+1)\times(L-m+1)$ matrix in equations~(\ref{blockdia})--(\ref{kernel3j}).
We may either implement the Wigner 3-$j$ expression~(\ref{kernel3j}) for
the elements~$D_{ll'}$, or use Gauss-Legendre
quadrature~\cite{Press+92} to evaluate the defining
integral~(\ref{blockdia}): 
\be
D_{ll'}=\int_{\cos\Theta}^1
X_{lm}(\arccos\mu)X_{l'm}(\arccos\mu)\,d\mu
\approx\sum_{j=1}^Jw_jX_{lm}(\arccos\mu_j) X_{l'm}(\arccos\mu_j),
\label{firstgauss}
\ee 
where $\mu_1,\mu_2,\ldots,\mu_J$ are roots of the Legendre
polynomial $P_J(\bar{\mu})$, rescaled from $-1\le\bar{\mu}_j\le 1$ to
$\cos\Theta\le\mu_j\le 1$, and
\mbox{$w_j=2(1-\bar{\mu}_j^2)^{-1}[P_J^{\prime}(\bar{\mu}_j)]^{-2}$,
with $j=1,2,\ldots,J$} are the associated integration weights. Only
the uppermost triangular matrix elements $D_{ll'},l\le l'$ are
computed explicitly; the lowermost elements are infilled using the
symmetry $D_{ll'}=D_{l'l}$.  The order of the Gauss-Legendre
integration is adjusted upward until the $L-m+1$ spatial-domain
eigenfunctions $g_1(\theta),g_2(\theta),\ldots,g_{L-m+1}(\theta)$
satisfy the orthogonality relations 
\be
2\pi\int_0^{\pi}g_{\alpha}g_{\beta}\sin\theta\,d\theta=\delta_{\alpha\beta}
\also 2\pi\int_0^{\Theta}g_{\alpha}g_{\beta}\sin\theta\,d\theta
=\lambda_{\alpha}\delta_{\alpha\beta}.
\label{fixedmortho}
\ee
to within machine
precision.  The same high-order Gauss-Legendre quadrature rule is used
to evaluate the orthogonality integrals.  The Legendre functions
$X_{lm}(\theta)$ are computed with high accuracy to very high degree
($l\approx 500$) using a recursive algorithm
\cite{Libbrecht85,Masters+98}. 

The second method is by solving the fixed-order version of the
Fredholm equation~(\ref{firsttimeint}) via the Nystrom method
\cite{Press+92}. Discretizing this equation by Gauss-Legendre
quadrature we obtain
\begin{equation}
\sum_{j'=1}^{J}w_{j'}D(\mu_j,\mu^{\prime}_{j'})g(\mu^{\prime}_{j'})=
\lambda\hspace{0.05em}g(\mu_j),\quad j=1,2,\ldots,J.
\label{interpol}
\end{equation}
Equation~(\ref{interpol}) can be rewritten as a symmetric
algebraic eigenvalue equation,
\begin{equation}
(\sfW\tilde{\sfD}\sfW)(\sfW\tilde{\textsf g})=
\lambda\hspace{0.05em}(\sfW\tilde{\textsf g}),
\label{wonehalf}
\end{equation}
where $\tilde{\textsf g}$ is a $J$-dimensional column vector with
entries $\tilde{g}_j=g(\mu_j)$, and where $\tilde{\sfD}$ and $\sfW$
denote the $J\times J$ matrices with elements
\mbox{$\tilde{D}_{jj'}=D(\mu_j,\mu^{\prime}_{j'})$ and
$W_{jj'}=\sqrt{w_j}\,\delta_{jj'}$.}  The eigenvalues $\lambda$ and
transformed eigenvectors $\sfW\tilde{\textsf g}$ are computed by
numerical diagonalization of the matrix $\sfW\tilde{\sfD}\sfW$.  The
order of integration $J$ is again chosen to ensure accurate
orthogonality of the spatial-domain eigenfunctions
$g_1(\theta),g_2(\theta),\ldots,g_{L-m+1}(\theta)$. In the zonal
($m=0$) case the choice $J=L+1$ renders both of the
integrations~(\ref{firstgauss}) and~(\ref{interpol}) exact; for
$m\not= 0$ we use a conservative, larger integration order $J$, since
the integrands are no longer polynomials.

Even for moderate values of the bandwidth $L$ and cap radius $\Theta$,
the smallest eigenvalues $\ldots,\lambda_{L-m},\lambda_{L-m+1}$ fall
below machine precision.  The associated, least well concentrated
eigenfunctions computed using either of the above two direct methods
are in that case essentially arbitrary orthogonal members of a
numerically degenerate eigenspace, and are no longer
accurate~\cite{Albertella+99}.  Because of this, it is not possible to
find the optimally excluded eigenfunctions of a small polar cap, or
equivalently the optimally concentrated eigenfunctions of a large cap,
by matrix diagonalization of~(\ref{firstgauss}). Fortunately, this
difficulty can be overcome by the third method, which is numerical
diagonalization of the tridiagonal Gr\"{u}nbaum
matrix~(\ref{gdefi1}). The roughly equant spacing of
their eigenvalues enables all of the associated eigenfunctions to be
calculated to within machine precision.  The spatiospectral
concentration factors $\lambda_1,\lambda_2,\ldots,\lambda_{L-m+1}$ are
computed to the same precision, either by {\it a posteriori\/} matrix
multiplication via~(\ref{fulleigen1}), or by Gauss-Legendre
integration of the orthogonality relation~(\ref{fixedmortho}).
Both the significant and the insignificant eigenvalues computed using
each of the above methods agree to within machine precision, providing
a useful numerical check. Diagonalization of the tridiagonal
matrix~(\ref{gdefi1}) is the only numerically stable
way to solve the concentration problem for either a large polar cap or
a large bandwidth~$L$. By extension, it is even
possible to use this formalism to compute
spacelimited eigenfunctions that are in the null space
\cite{Miranian2004}.  The above results apply to the
double-cap case if using the appropriate Gr\"unbaum
matrix~(\ref{gdefi}). 

\subsection{Concentration within an arbitrarily shaped region}
\label{subsec:Comp:arbit}
We solve the spatiospectral concentration problem for an arbitrarily shaped
region $R$ by numerical diagonalization of the $\Lpot\times\Lpot$
matrix with elements $D_{lm,l'm'}$
defined by equation~(\ref{Dlmlmpdef}).
Given a (splined) boundary of~$R$,
we first find the northernmost and southernmost points,
with colatitudes $\theta_{\rm n}$ and $\theta_{\rm s}$.
For every $\theta_{\rm n}\le\theta\le\theta_{\rm s}$,
we then find the easternmost and westernmost points,
with longitudes $\phi_{\rm e}(\theta)$ and $\phi_{\rm w}(\theta)$.
In the case of a non-convex region with indentations and protuberances,
there may be several such eastern and western boundary points,
which we shall index with an additional subscript $i=1,2,\ldots,I$.
The integral over longitude,
\begin{equation}
\Phi_{mm'}(\theta)=\sum_{i=1}^I\int_{\phi_{{\rm w}i}}
^{\phi_{{\rm e}i}}\!\left\{\!\!\begin{array}{c} \cos m\phi \\ \sin
m\phi\end{array}\!\!\right\} 
\left\{\!\!\begin{array}{c} \cos m'\phi \\ \sin
m'\phi\end{array}\!\!\right\}d\phi, 
\label{easyone}
\end{equation}
is done analytically, and we use Gauss-Legendre quadrature to
compute the remaining integral over colatitude:
\ber
D_{lm,l'm'}&=&\int_{\mu_{\rm n}}^{\mu_{\rm s}}
X_{lm}(\arccos\mu)X_{l'm'}(\arccos\mu)\Phi_{mm'}(\arccos\mu)\,d\mu\nnr\\
&\approx&\sum_{j=1}^Jw_jX_{lm}(\arccos\mu_j)X_{l'm'}(\arccos\mu_j)
\Phi_{mm'}(\arccos\mu_j).
\label{hardone}
\eer
As in the case of a polar cap, we adjust the order of the integration
$J$ upward until the spatial-domain eigenfunctions
$g_1(\rhat),g_2(\rhat),\ldots,g_{\Lpot}(\rhat)$
satisfy the orthogonality relations
\be
\into g_{\alpha}g_{\beta}\domg=\delta_{\alpha\beta},\also
\intr  g_{\alpha}g_{\beta}\domg=\lambda_{\alpha}\delta_{\alpha\beta}.
\label{orthog}
\ee
to within machine precision. There is no analogue to
Gr\"{u}nbaum's procedure for an arbitrarily shaped region, so only the
eigenfunctions associated with eigenvalues that are above machine
precision can be computed accurately. In most practical
applications\cite{Simons+2006b}, this is not a limitation, since we
are generally interested only in the computable, well concentrated
eigenfunctions $g_1(\rhat),g_2(\rhat),\ldots,g_N(\rhat)$, which are
associated with the numerically significant eigenvalues 
$\lambda_1,\lambda_2,\ldots,\lambda_N$ where $N$ is the Shannon
number~(\ref{tracedef}). 

\subsection{Concentration within a non-polar circular cap}
\label{subsec:Comp:away}
One of the principal applications of spherical Slepian functions in
geophysics and planetary
physics\cite{Kido+2003,McGovern+2002,Simons+97b,Simons+97a} is
to analyze measurements within a circularly symmetric region centered
upon an arbitrary geographical location $\theta_0,\phi_0$.  The
preferred procedure for determining the required optimally
concentrated eigenfunctions is first to compute the spherical harmonic
coefficients $g_{lm}$ of the eigenfunctions concentrated within a
polar cap $0\leq\theta\leq\Theta$, and then to rotate these to the
desired cap location
\cite{Blanco+97,Dahlen+98,Edmonds96,Masters+98}. The actual windowing
of the data for further analysis may either be carried out in the
spectral domain~\cite{Simons+97a}, or, more simply, by straightforward
multiplication after transformation of the rotated eigenfunctions to
the spatial domain.  If one wishes to avoid spherical harmonic
rotation, it is also possible to compute the rotated eigenfunctions
directly, by performing the numerical integration in
equation~(\ref{easyone}) on the analytically prescribed boundary of a
cap of radius $\Theta$ centered at $\theta_0,\phi_0$, given by
\begin{equation}
\phi_{\rm w,e}(\theta)=\phi_0\mp\Delta\phi(\theta) \quad \mbox{where} \quad
\Delta\phi(\theta)=\frac{\arccos(\cos\Theta-\cos\theta\cos\theta_0)}
{\sin\theta\sin\theta_0}.
\end{equation}

\section{CONCLUSIONS}

Spherical Slepian functions provide a natural solution to the problem
of having a polar gap in the satellite coverage of planetary
gravitational or magnetic fields. Indeed, the ill-posed estimation
problem of finding a source-level potential from noisy observations
taken at an altitude over an incomplete region of coverage has natural
connections to Slepian's spherical problem of spatiospectral
localization. We have proposed a method~\cite{Simons+2006b}
that expands the source field in terms of a truncated basis set of
spherical Slepian functions, and compared its statistical performance
with the damped least-squares method in the spherical
harmonic basis. The optimally truncated Slepian method performs nearly
as well as the optimally damped spherical harmonic method, but it has
the significant advantage of an intuitive separation of the estimation
bias and variance over those Slepian functions sensitive to the
uncovered and covered regions, respectively. In this contribution we
have tied up some loose ends in our published work on spherical
Slepian functions by illustrating their asymptotic self-similarity in
a flat-Earth limit and by focusing attention on numerical
implementation issues not published elsewhere. 

\acknowledgments

Financial support for this work has been provided by the
U.~S.~National Science Foundation under Grants EAR-0105387 awarded to
FAD and EAR-0710860 to FJS, and by a U.~K.~Natural Environmental
Research Council New Investigator Award (NE/D521449/1) and a Nuffield
Foundation Grant for Newly Appointed Lecturers (NAL/01087/G) awarded
to FJS at University College London. FJS thanks Mark Wieczorek for
many fruitful discussions and the D{\'e}partement de G{\'e}ophysique
Spatiale et Plan{\'e}taire at the Institut de Physique du Globe de
Paris for its hospitality and financial support. Computer algorithms
are made available on \url{www.frederik.net}.

\newpage 

\bibliography{/home/fjsimons/BIBLIO/bib}

\begin{thebibliography}{10}

\bibitem{Slepian+60}
D.~Slepian and H.~O. Pollak, ``Prolate spheroidal wave functions, {F}ourier
  analysis and uncertainty --- {I},'' {\em Bell Syst.~Tech.~J.}~{\bf 40}(1),
  pp.~43--63, 1960.

\bibitem{Thebault+2006}
E.~Th{\'e}bault, J.~J. Schott, and M.~Mandea, ``Revised spherical cap harmonic
  analysis ({R-SCHA}): Validation and properties,'' {\em J.~Geophys.~Res.}~{\bf
  111}(B1), pp.~B01102, doi:10.1029/2005JB003836, 2006.

\bibitem{Lesur2006}
V.~Lesur, ``Introducing localized constraints in global geomagnetic field
  modelling,'' {\em Earth Planets Space}~{\bf 58}(4), pp.~477--483, 2006.

\bibitem{Albertella+99}
A.~Albertella, F.~Sans{\`o}, and N.~Sneeuw, ``Band-limited functions on a
  bounded spherical domain: the {S}lepian problem on the sphere,'' {\em
  J.~Geodesy}~{\bf 73}, pp.~436--447, 1999.

\bibitem{Simons+2006b}
F.~J. Simons and F.~A. Dahlen, ``Spherical {S}lepian functions and the polar
  gap in geodesy,'' {\em Geophys.~J.~Int.}~{\bf 166}, pp.~1039--1061,
  doi:10.1111/j.1365--246X.2006.03065.x, 2006.

\bibitem{Holschneider+2003}
M.~Holschneider, A.~Chambodut, and M.~Mandea, ``From global to regional
  analysis of the magnetic field on the sphere using wavelet frames,'' {\em
  Phys.~Earth Planet.~Inter.}~{\bf 135}, pp.~107--124, 2003.

\bibitem{Freeden+2004}
W.~Freeden and V.~Michel, ``Orthogonal zonal, tesseral and sectorial wavelets
  on the sphere for the analysis of satellite data,'' {\em
  Adv.~Comput.~Math.}~{\bf 21}(1--2), pp.~181--217, 2004.

\bibitem{Slepian83}
D.~Slepian, ``Some comments on {F}ourier analysis, uncertainty and modeling,''
  {\em {SIAM} {R}ev.}~{\bf 25}(3), pp.~379--393, 1983.

\bibitem{Wieczorek+2005}
M.~A. Wieczorek and F.~J. Simons, ``Localized spectral analysis on the
  sphere,'' {\em Geophys.~J.~Int.}~{\bf 162}(3), pp.~655--675,
  doi:10.1111/j.1365--246X.2005.02687.x, 2005.

\bibitem{Simons+2006a}
F.~J. Simons, F.~A. Dahlen, and M.~A. Wieczorek, ``Spatiospectral concentration
  on a sphere,'' {\em {SIAM} {R}ev.}~{\bf 48}(3), pp.~504--536,
  doi:10.1137/S0036144504445765, 2006.

\bibitem{Pail+2001}
R.~Pail, G.~Plank, and W.-D. Schuh, ``Spatially restricted data distributions
  on the sphere: the method of orthonormalized functions and applications,''
  {\em J.~Geodesy}~{\bf 75}, pp.~44--56, 2001.

\bibitem{Gilbert+77}
E.~N. Gilbert and D.~Slepian, ``Doubly orthogonal concentrated polynomials,''
  {\em {SIAM} J.~Math.~Anal.}~{\bf 8}(2), pp.~290--319, 1977.

\bibitem{Grunbaum+82}
F.~A. Gr{\"u}nbaum, L.~Longhi, and M.~Perlstadt, ``Differential operators
  commuting with finite convolution integral operators: some non-{A}belian
  examples,'' {\em {SIAM} J.~Appl.~Math.}~{\bf 42}(5), pp.~941--955, 1982.

\bibitem{Percival+93}
D.~B. Percival and A.~T. Walden, {\em Spectral Analysis {f}or Physical
  Applications, Multitaper {a}nd Conventional Univariate Techniques}, Cambridge
  Univ.~Press, New York, 1993.

\bibitem{Xu1992b}
P.~Xu, ``The value of minimum norm estimation of geopotential fields,'' {\em
  Geophys.~J.~Int.}~{\bf 111}, pp.~170--178, 1992.

\bibitem{Holme+95}
R.~Holme and J.~Bloxham, ``Alleviation of the {B}ackus effect in geomagnetic
  field modelling,'' {\em Geophys.~Res.~Lett.}~{\bf 22}(13), pp.~1641--1644,
  1995.

\bibitem{Mallat98}
S.~Mallat, {\em A Wavelet Tour {o}f Signal Processing}, Academic Press, San
  Diego, Calif., 1998.

\bibitem{Lowes74}
F.~J. Lowes, ``Spatial power spectrum of the main geomagnetic field and
  extrapolation to core,'' {\em Geophys.~J.~R.~Astron.~Soc.}~{\bf 36}(3),
  pp.~717--730, 1974.

\bibitem{Gubbins83}
D.~Gubbins, ``Geomagnetic field analysis -- {I}. {S}tochastic inversion,'' {\em
  Geophys.~J.~R.~Astron.~Soc.}~{\bf 73}(3), pp.~641--652, 1983.

\bibitem{Tegmark96a}
M.~Tegmark, ``A method for extracting maximum resolution power spectra from
  microwave sky maps,'' {\em Mon.~Not.~R.~Astron.~Soc}~{\bf 280}, pp.~299--308,
  1996.

\bibitem{Peebles73}
P.~J.~E. Peebles, ``Statistical analysis of catalogs of extragalactic objects.
  {I}. {T}heory,'' {\em Astroph.~J.}~{\bf 185}, pp.~413--440, 1973.

\bibitem{Tegmark95}
M.~Tegmark, ``A method for extracting maximum resolution power spectra from
  galaxy surveys,'' {\em Astroph.~J.}~{\bf 455}, pp.~429--438, 1995.

\bibitem{Dahlen+98}
F.~A. Dahlen and J.~Tromp, {\em Theoretical Global Seismology}, Princeton
  Univ.~Press, Princeton, N.~J., 1998.

\bibitem{Edmonds96}
A.~R. Edmonds, {\em Angular Momentum in Quantum Mechanics}, Princeton
  Univ.~Press, Princeton, N.J., 1996.

\bibitem{Kanwal71}
R.~P. Kanwal, {\em Linear Integral Equations; Theory {a}nd Technique}, Academic
  Press, New York, 1971.

\bibitem{Tricomi70}
F.~G. Tricomi, {\em Integral Equations}, Interscience, New York, 5~ed., 1970.

\bibitem{Messiah2000}
A.~Messiah, {\em Quantum Mechanics}, Dover, New York, 2000.

\bibitem{Landau65}
H.~J. Landau, ``On the eigenvalue behavior of certain convolution equations,''
  {\em Trans.~Am.~Math.~Soc.}~{\bf 115}, pp.~242--256, 1965.

\bibitem{Slepian+65}
D.~Slepian and E.~Sonnenblick, ``Eigenvalues associated with prolate spheroidal
  wave functions of zero order,'' {\em Bell Syst.~Tech.~J.}~{\bf 44}(8),
  pp.~1745--1759, 1965.

\bibitem{Slepian64}
D.~Slepian, ``Prolate spheroidal wave functions, {F}ourier analysis and
  uncertainty --- {IV}: {E}xtensions to many dimensions; generalized prolate
  spheroidal functions,'' {\em Bell Syst.~Tech.~J.}~{\bf 43}(6),
  pp.~3009--3057, 1964.

\bibitem{Amado+85}
R.~D. Amado, K.~Stricker-Bauer, and D.~A. Sparrow, ``Semiclassical methods and
  the summation of the scattering partial wave series,'' {\em Phys.~Rev.\
  C}~{\bf 32}(1), pp.~329--332, 1985.

\bibitem{Dahlen80}
F.~A. Dahlen, ``A uniformly valid asymptotic representation of normal mode
  multiplet spectra on a laterally heterogeneous {E}arth,'' {\em
  Geophys.~J.~R.~Astron.~Soc.}~{\bf 62}(2), pp.~225--247, 1980.

\bibitem{Hilb19}
E.~Hilb, ``{\"U}ber die {L}aplacesche {R}eihe,'' {\em Math.~Z.}~{\bf 5}, p.~17,
  1919.

\bibitem{Szego75}
G.~Szeg{\"o}, {\em Orthogonal Polynomials}, American Mathematical Society,
  Providence, R.I., 4~ed., 1975.

\bibitem{Olver97}
F.~W.~J. Olver, {\em Asymptotics {a}nd Special Functions}, A. K. Peters,
  Wellesley, Mass., 1997.

\bibitem{Jeffreys+88}
H.~Jeffreys and B.~S. Jeffreys, {\em Methods of Mathematical Physics},
  Cambridge Univ.~Press, Cambridge, UK, 3~ed., 1988.

\bibitem{Press+92}
W.~H. Press, S.~A. Teukolsky, W.~T. Vetterling, and B.~P. Flannery, {\em
  Numerical Recipes {i}n {FORTRAN}: The Art {o}f Scientific Computing},
  Cambridge Univ.~Press, 2nd~ed., 1992.

\bibitem{Libbrecht85}
K.~G. Libbrecht, ``Practical considerations for the generation of large-order
  spherical harmonics,'' {\em Solar Physics}~{\bf 99}(1--2), pp.~371--373,
  1985.

\bibitem{Masters+98}
G.~Masters and K.~Richards-Dinger, ``On the efficient calculation of ordinary
  and generalized spherical harmonics,'' {\em Geophys.~J.~Int.}~{\bf 135}(1),
  pp.~307--309, 1998.

\bibitem{Sneeuw94}
N.~Sneeuw, ``Global spherical harmonic-analysis by least-squares and numerical
  quadrature methods in historical perspective,'' {\em Geophys.~J.~Int.}~{\bf
  118}(3), pp.~707--716, 1994.

\bibitem{Miranian2004}
L.~Miranian, ``Slepian functions on the sphere, generalized {G}aussian
  quadrature rule,'' {\em Inv.~Prob.}~{\bf 20}, pp.~877--892, 2004.

\bibitem{Kido+2003}
M.~Kido, D.~A. Yuen, and A.~P. Vincent, ``Continuous wavelet-like filter for a
  spherical surface and its application to localized admittance function on
  {M}ars,'' {\em Phys.~Earth Planet.~Inter.}~{\bf 135}, pp.~1--14, 2003.

\bibitem{McGovern+2002}
P.~J. McGovern, S.~C. Solomon, D.~E. Smith, M.~T. Zuber, M.~Simons, M.~A.
  Wieczorek, R.~J. Phillips, G.~A. Neumann, O.~Aharonson, and J.~W. Head,
  ``Localized gravity/topography admittance and correlation spectra on {M}ars:
  {I}mplications for regional and global evolution,'' {\em
  J.~Geophys.~Res.}~{\bf 107}(E12), pp.~5136, doi:10.1029/2002JE001854, 2002.

\bibitem{Simons+97b}
M.~Simons and B.~H. Hager, ``Localization of the gravity field and the
  signature of glacial rebound,'' {\em Nature}~{\bf 390}, pp.~500--504,
  December 1997.

\bibitem{Simons+97a}
M.~Simons, S.~C. Solomon, and B.~H. Hager, ``Localization of gravity and
  topography: Constraints on the tectonics and mantle dynamics of {V}enus,''
  {\em Geophys.~J.~Int.}~{\bf 131}, pp.~24--44, October 1997.

\bibitem{Blanco+97}
M.~A. Blanco, M.~Fl{\'o}rez, and M.~Bermejo, ``Evaluation of the rotation
  matrices in the basis of real spherical harmonics,'' {\em J.~Mol.~Struct.
  (Theochem)}~{\bf 419}, pp.~19--27, 1997.

\end{thebibliography}
\bibliographystyle{spiebib}

\newpage

\end{document}